\documentstyle[aps,prl,epsf]{revtex}
\title{\strut{\vadjust{\vskip-0.5in\strut\hfill{\normalsize
	IASSNS-HEP-99/100\vskip0.25in}}} 
Charged stripes from alternating static magnetic field}
       
\author{
Oleg Tchernyshyov\cite{e-mail-Oleg} and Frank Wilczek\cite{e-mail-Frank}}
\address{School of Natural Sciences, 
Institute for Advanced Study, 
Princeton, New Jersey 08540 }

\begin{document}
\twocolumn[\hsize\textwidth\columnwidth\hsize\csname %
  @twocolumnfalse\endcsname
\maketitle

\begin{abstract}
We motivate and perform a calculation of the energy of a
cold fluid of charged fermions in the presence of a striped magnetic
background.  We find that a non-trivial value for the doping density
on the walls is preferred.
\end{abstract}
\bigskip\bigskip
]

The concept of ``fictitious'' or ``statistical'' gauge fields has
proved quite fruitful in the study of the quantum Hall effect.  Gauge
fields have 
also figured prominently in theoretical attempts to understand other
highly correlated electron systems in two dimensions, especially the
doped cuprate planes famous for supporting high-temperature
superconductivity ~\cite{FSAS,LeeWen,BalentsFN}.  
In the mean-field treatment of such theories, one
is often involved in expanding around non-zero background values for
the fictitious magnetic field.  This background field spontaneously
breaks P and T symmetry, and is therefore naturally associated with a
$Z_2$ order parameter.  Unfortunately, experiments to detect
macroscopic P and T violation in real materials have produced negative
results.

On the other hand, recently a number of experiments have discovered
evidence for inhomogeneous electronic structure in the doped cuprate
planes ~\cite{Tranquada,B1,B2,Imai}.  
Existing evidence is consistent with the idea that this
structure is roughly describable in terms of antiferromagnetic domains
separated by narrow antiphase walls, with the dopants concentrated on
the walls, at least for small doping.  In view of this observed
inhomogeneity, it is interesting and natural to consider the
possibility of stripe structure in the context of fictitious gauge
fields.  Indeed, this concept has several advantages that are evident
prior to any detailed calculation.  The empirical difficulty of
macroscopic P and T violation is ameliorated, and the $Z_2$ nature of
the order provides a {\it raison d'etre\/} for stable domain walls.

On deeper consideration it appears that there can be, in the context
of these ideas, a compelling connection between doping and stripe
structure.  Indeed, let us suppose that the preferred bulk phase,
realized in the absence of doping, prefers values of the magnetic
field $\pm B$, with one sign chosen uniformly throughout the plane, at
which the electrons just fill an integer number of Landau levels.  As
one dopes away from this preferred filling, two things might happen.
It might be that one just has to live with an unfilled band.
Alternatively, it might be preferable to produce regions with both
$\pm B$, separated by narrow walls.  One could have then have the
filled band in bulk, accommodating dopants on the domain walls.

To determine the viability of this concept, we have performed the
following calculation.

Consider ideal fermions in 2 dimensions in uniform magnetic field $B$.
Initially, exactly $N$ first Landau levels 
($n=0,1,\ldots,N-1$) are filled.  The energy
density is ${\cal E}_0 = \frac{1}{2}N\hbar\omega_c\,\rho$, where 
$\rho = NeB/2\pi\hbar c$ is the fermion density and $\omega_c=eB/mc$ is 
the cyclotron frequency.  

Add or remove $\Delta\rho$ fermions per unit area.  If the magnetic
field stays commensurate with density, keeping $N$ filled Landau levels,
$\Delta {\cal E} = N\hbar\omega_c\,\Delta\rho$.  Otherwise, if $B$
remains fixed,
\begin{equation}
\Delta {\cal E} = 
\left\{
\begin{array}{ll}
(N+1/2) \hbar\omega_c\,\Delta\rho & \mbox {if } \Delta\rho>0,\\
(N-1/2) \hbar\omega_c\,\Delta\rho & \mbox {if } \Delta\rho<0.
\end{array}
\right.
\label{dE-uniform}
\end{equation}

\begin{figure}
\begin{center}
\leavevmode
\epsfxsize \columnwidth
\epsffile{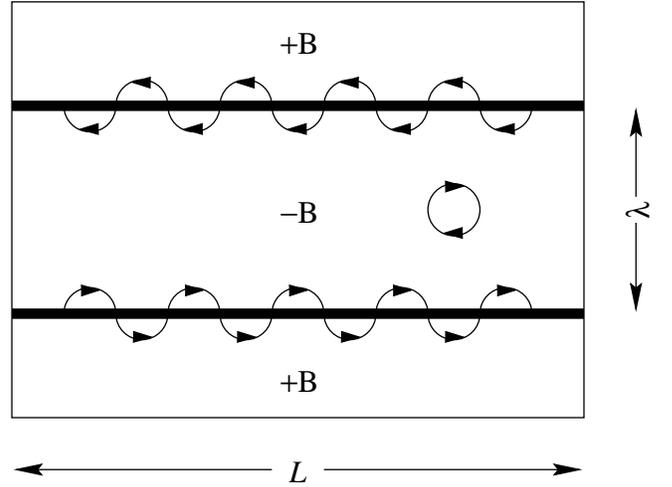}
\end{center}
\caption{Striped arrangement of magnetic field.  Also shown are the
semi-classical electron orbits.  Notice that states on the wall have a
preferred direction of momentum.}
\label{geometry}
\end{figure}

In our alternative scenario, magnetic field is allowed to
alternate between $B$ and $-B$, thus producing magnetic domains 
(Fig. \ref{geometry}).  Landau
levels are distorted in the vicinity of domain walls.  Midgap states
induced by the domain wall accommodate newly doped fermions (or holes),
so that doped charges form stripes along the domain walls.  We wish to
compare the energy of this state
of a doped system with that in a fixed uniform
magnetic field. 

The linear density of charge on a stripe $\nu$ and the stripe spacing
$\lambda$ are related to the average density of added charge according
to 
$\Delta\rho = \nu/\lambda$.  For small $\Delta\rho$, stripes are far
apart compared to the width of the wave functions, 
and they can be considered independently of one another.  In units of
magnetic length $(\hbar c/eB)^{1/2}$, this requires $\lambda \gg
(N-1/2)^{1/2}$.

For a single domain wall of length $L$ with ${\cal N}=\nu L$ added
particles, the energy is $E({\cal N}) = E_0 + \epsilon(\nu)L$, where
$E_0$ is the energy of the undoped system {\em without\/} the domain
wall and $\epsilon(\nu)$ is the energy density of a domain wall (per
unit length).  The relative energy of the system per unit area is
\begin{equation}
\Delta{\cal E} = \frac{\epsilon(\nu)}{\lambda} 
= \frac{\epsilon(\nu)}{\nu}\Delta\rho
= \frac{E({\cal N})-E_0}{\cal N}\Delta\rho.
\label{dE-stripes}
\end{equation}
Thus the preferred linear charge density $\nu_0$ is found by minimizing 
$\epsilon(\nu)/|\nu|$, the energy per doped particle (or hole).  
Formation of stripes is advantageous if this energy is lower than 
that in a uniform field $B$:
\begin{equation}
\frac{\epsilon(\nu_0)}{|\nu_0|} < 
\left\{
\begin{array}{rl}
(N+1/2) \hbar\omega_c & \mbox {for electrons},\\
-(N-1/2) \hbar\omega_c & \mbox {for holes.} 
\end{array}
\right.
\label{stripes-condition}
\end{equation}

{\em Energy spectrum of a single domain wall}:  For a stripe along the
$x$ axis, $B(y) = -B\,{\rm sgn}(y)$.  It is convenient to work in the
Landau gauge $A_x = B|y|$, $A_y = 0$, so that translation symmetry in
$x$ is manifest.   Wavefunctions
$\psi(y)\exp{(ikx)}$ then satisfy the Schroedinger equation
\begin{equation}
-\psi''(y) + (|y|-k)^2\, \psi(y) = 2E\, \psi(y), 
\label{Lifshits}
\end{equation}
in appropriate units of length and energy.

As a function of momentum $k$
along the stripe, there are 3 simple regimes.  
If $k > 0$ and
large one has two well-separated
parabolic wells.  Then there are 
doublets near the levels of the harmonic oscillator, with an
exponentially small splitting between the symmetric and antisymmetric
states.  If $k < 0$ and large one has a single wedge-like well.  Energy levels
start (at least) at $|k|$ and thus do not contribute to low Landau
levels.  Finally if $k=0$, one has the simple harmonic oscillator.  

Thus as $k$ varies from
$+\infty$ to 0, the $n$-th Landau level splits into symmetric and
antisymmetric branches.  The energy of the symmetric branch $E_+(k)$
is {\em lower} than $n+1/2$ for large k.  At $k=0$, $E_+ = 2n+1/2$ and
$E_- = 2n+3/2$.

\begin{figure}
\begin{center}
\leavevmode
\epsfxsize \columnwidth
\epsffile{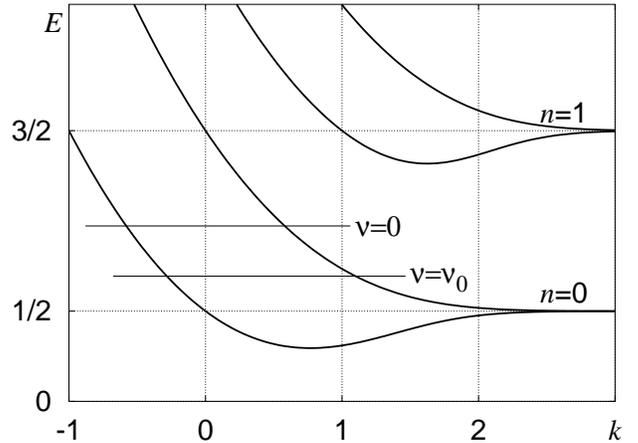}
\end{center}
\caption{Energies of the symmetric and antisymmetric eigenstates of a
stripe, as functions of k.  Straight lines are the Fermi levels for 0
and optimal dopant density on a stripe.}
\label{spectrum}
\end{figure}

Solutions of the Schroedinger equation (\ref{Lifshits}) 
regular at $y=\infty$ are parabolic cylinder functions \cite{WW}
\begin{equation}
\psi(y) = 
\left\{
\begin{array}{cl}
D_{E-1/2}\left((y-k)\sqrt{2}\right) & \mbox{ if } y \geq 0, \\
\pm\psi(|y|) & \mbox{ if } y < 0.
\end{array}
\right.
\label{parabolic-cylinder}
\end{equation}
The spectra $E_\pm(k)$ are determined by a boundary condition at
$y=0$: $\psi'(+0)=0$ for symmetric states and $\psi(+0)=0$ for
antisymmetric ones.  All the necessary calculations are now fast
computations for MAPLE.  The results are shown in Fig. \ref{spectrum}. 

{\em Energy of a single stripe}: Further analysis of the energy
competition is greatly simplified by the existence of a one-to-one
correspondence between states with and without a domain wall.  Since
$k$ is quantized in units $\frac{2\pi}{L}$ in both cases, there are no
phase shifts to worry about.  We will consider in detail the case
of $N=1$ filled Landau level.

In uniform magnetic field, all states of the lowest Landau level
($n=0$, $-\infty<k<+\infty$) are occupied.  To obtain the same {\em
number of particles\/} in the system with a single domain wall in the
most economical way, we must partially fill the two lowest branches
($n=0$), as follows.  First fill all the symmetric states with $k \geq
0$ and all antisymmetric states with $k > 0$; then slightly rearrange
occupied and empty states within the two bands to achieve the state
with lowest energy.  The Fermi energy and momentum are determined by
the equation $E_+(-k_F) = E_-(k_F) = E_F$.  Note that the Fermi level
lies below the bottom of the symmetric $n=1$ band (Figure
\ref{spectrum}).

Without doping, a stripe has a positive linear density of energy.  At
large doping levels, however, $E({\cal N})-E_0 \sim -{\cal N}/2 + C$, as holes
are doped into unperturbed Landau states away from the domain wall.
The constant $C$ is {\em negative}.  This becomes evident from Figure
2 if one considers
$E_F$ just above $1/2$, for then most of the remaining electrons are in the
symmetric band, {\it i.e.}, below $1/2$.  The energy per doped
hole is
\begin{equation}
\frac{\epsilon(\nu)}{|\nu|}
= \frac{E({\cal N})-E_0}{|{\cal N}|} \sim -\frac{1}{2} + \frac{c}{|\nu|}, 
\hskip 5mm 
c = C/L < 0.
\end{equation}
Because $\epsilon(\nu)/|\nu|>0$ for $\nu\to -0$, this function has a
minimum at a finite $\nu_0$.  Moreover, condition
(\ref{stripes-condition}) is satisfied.  Thus for $N=1$ the stripe
phase has lower energy than the phase with a fixed uniform magnetic
field, at least for sufficiently low doping levels, such that stripe
wavefunctions do not overlap.

\begin{figure}
\begin{center}
\leavevmode
\epsfxsize \columnwidth
\epsffile{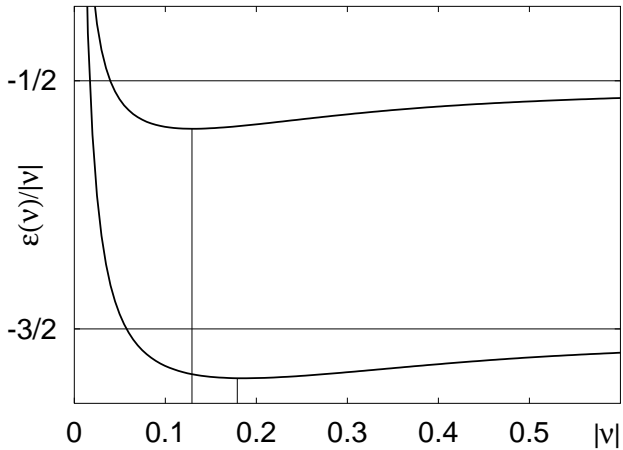}
\end{center}
\caption{Energy per added hole in the stipe phase, for $N=1$ and 2 
filled Landau levels.  Both curves display shallow minima at values
below the corresponding no-stripe values.}
\label{energy}
\end{figure}

In Figure 3, we have displayed
the energy per added hole $\epsilon(\nu)/|\nu|$ for $N=1$ and
2 filled Landau levels in bulk.  
For $N=1$, we find $|\nu_0| \approx 0.13$ particles
per magnetic length.

On a lattice with flux $\phi$ piercing an elementary plaquette, the
lattice constant is $a = \phi^{1\over 2}$.  The preferred doping density 
in lattice units is therefore
\begin{equation}
|\nu_0| \approx \frac{0.13 \mbox{ particles}}{\mbox{magn. length}}
= \frac{0.13\, \phi^{1/2} \mbox{ particles}}{\mbox{lattice constant}}.
\end{equation}
%** references for flux value
If $\phi = \pi$, as suggested theoretically
~\cite{Affleck,Wiegmann,Lieb,FSAS},
we find 
$|\nu_0|\approx 0.23$ particles per lattice constant.

%\section{   }
{\em Discussion}:
In our picture, the domain walls resemble quantum Hall edges.  They
support currents, alternating in direction from wall to wall, in the
ground state.

Our calculation has been highly idealized, of course.  For one thing,
we have ignored any possibility of energy intrinsically associated
with the fictitious field (other than the implicit constraint
enforcing values $\pm B$).  While this approximation is in the spirit
of fictitious Chern-Simons fields, or of fields implementing
constraints, which have no independent dynamics, in the absence of a
detailed microscopic model we cannot assess its accuracy.  Similarly,
we have been shamelessly opportunistic in maneuvering back and forth
between lattice and continuum.  With all due reserve, it nevertheless
seems appropriate to mention that the sort of model discussed here is
remarkable and virtually unique, as far as we are aware, in predicting
a non-trivial preferred dopant density along the stripes.  Moreover,
the preferred numerical value, which has emerged from a dynamical
calculation containing no disposable continuous parameters, is
consistent with the observed one.

{\em Acknowledgments:} We thank M.~M. Fogler and A. Zee for helpful 
discussions.  Research supported in part by DOE grant
DE-FG02-90ER40542.

\end{document}